# A new approach for measurement of $Cr^{4+}$ concentration in $Cr^{4+}$:YAG transparent materials: some conceptual difficulties and possible solutions

M. Chaika[1], R. Lisiecki[1], K. Lesniewska-Matys[2], O.M. Vovk[3].

[1]Institute of Low Temperature and Structure Research Polish Academy of Science, Okolna 2, 50422, Wroclaw, Poland

[2]Institute of Electronic Materials Technology, 133 Wolczynska Str., Warsaw 01-919, Poland

[3]Institute for Single Crystals, National Academy of Sciences of Ukraine, 60 Nauky Avenue, Kharkiv, 61072, Ukraine

**Abstract**

In the present work we provide an analysis of the accuracy of the calculation of $Cr^{4+}$ concentration in $Cr^{4+}$:YAG using absorption spectroscopy. We propose a new approach based on the convenient optical spectroscopy to estimate the $Cr^{4+}$ concentration in Cr:YAG using survey absorption spectra. The Smakula–Dexter formula is usually used for this purpose. However, the uncertainties in the values of oscillator strengths for $Cr^{4+}$ absorption bands and, moreover, in the deconvolution of $Cr^{4+}$ absorption spectra make it difficult to calculate the $Cr^{4+}$ concentrations with high accuracy. The following formulas were proposed for calculation of $Cr^{4+}$ concentration: $n_A = 1.2 \cdot 10^{17} H_{480}$, $n_D = 5.4 \cdot 10^{17} H_{1030}$, where $n_A$, cm$^{-3}$ and $n_D$, cm$^{-3}$ are calculated concentrations of octahedrally and tetrahedrally coordinated $Cr^{4+}$ ions, respectively, $H_{480}$ and $H_{1030}$ are absorption coefficients of $Cr^{4+}$ ions at 480 nm and 1030 nm, respectively. The actual concentration $n_0$ is in the range of $n/2 < n_0 < 2 \cdot n$, where n is calculated value.

## 1. Introduction

Due to simplicity, compactness, and low cost, passive Q-switchers is very attractive for solid-state lasers applications. Due to the fact that Nd:YAG is one of the most spread CW solid-state lasers, $Cr^{4+}$-doped garnets got a lot of attention in recent decades as passive Q-switches at ~ 1 μm [1]. Recently, a number of $Cr^{4+}$-doped materials have been proposed for this purpose, such as $Y_3Al_5O_{12}$, $Y_3Sc_2Ga_3O_{12}$, $Lu_3Al_5O_{12}$, and $Gd_3Ga_5O_{12}$ [2-4]. The Q-switching properties of these materials is based on saturation absorption of 1 μm light by $Cr^{4+}$ ions in tetrahedrally coordinated site [5,6]. Formation of tetravalent chromium ions requires co-doping by divalent impurity ions, however, the mechanism of formation of $Cr^{4+}$ ions is still unclear [7-10]. Moreover, the issue related to concentration of $Cr^{4+}$ ions remain under discussion.

The concentration of $Cr^{4+}$ ions can be calculated using the Avizonis-Grotbeck theory [11]. This model is valid for the case of a slow saturable absorber and takes into account excited state absorption in the studied material. However, the calculated values of the absorption cross-sections of $Cr^{4+}$:YAG differ in one order of magnitude in various papers [12-16] introducing the uncertainty in the calculation of $Cr^{4+}$ concentration. Moreover, this way is complicated that limits the applicability of this approach. On the other hand, the concentration of $Cr^{4+}$ ions can be calculated using the optical absorption spectroscopy according to the methodology described by Feldman et. al. [12]. However, such method has the same disadvantages as Avizonis-Grotbeck calculation requiring the oscillator strengths of $Cr^{4+}$ absorption bands, which can be calculated based on the results obtained from Avizonis-Grotbeck theory. As a result, the reported values of $Cr^{4+}$ concentration differ in one order of magnitude for materials with similar absorption.

This paper focuses on the calculation of $Cr^{4+}$ concentration in $Cr^{4+}$:YAG materials using optical spectroscopy methods. The problems and the pitfalls of the calculations of $Cr^{4+}$ concentration in YAG materials are analyzed. The simplified formula is proposed for calculation of $Cr^{4+}$ concentration from optical absorption spectra.

2. Experimental

The $Cr^{4+},Ca^{2+}$:YAG ceramics were synthesized in the Institute for Single Crystals, Ukraine [7,17-19]. Fig. 1 presents a photograph of the transparent $Cr^{4+},Ca^{2+}$:YAG ceramics. Absorbance spectra were measured using Perkin-Elmer Lambda-35 UV/Vis spectrophotometer (wavelength range: 190-1100 nm) at room temperature.

The nonlinear optical properties of transparent $Cr^{4+}$:YAG ceramics were evaluated by the methods of saturated absorption/transmission spectroscopy using Q-switched Nd:YAG laser Litron Nano S (1064 nm, the pump pulse duration was 7 ns). The measuring set consisted of two energy detectors, set of filters, and beam splitter. The thickness of the sample was 1 mm, it was polished but without AR-coating.

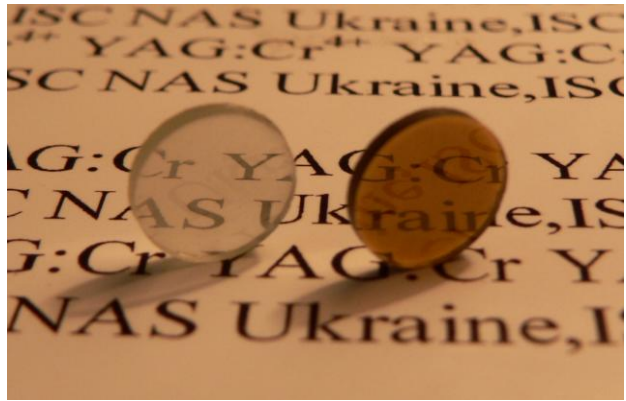

Fig. 1: Photograph of the $Cr^{4+},Ca^{2+}$:YAG ceramics.

## 3. Results

For the benefit of the reader, the present paper focuses only on the issues related to $Cr^{4+}$ content estimation. The microstructure and luminescence properties of $Cr^{4+}$:YAG ceramics of the same composition were fully described in our previous work [7,17-21]. Firstly, the mechanism of $Cr^{4+}$ formation is described for a better understanding of the experimental results.

The $Cr^{4+}$ ions in a YAG lattice can occupy both octahedral and tetrahedral positions, the composition of YAG can be written as $C_3A_2D_3O_{12}$, where dodecahedrally, octahedrally, and tetrahedrally oxygen coordinated cation sites denotes as "C", "A" and "D" respectively [22]. The $Al^{3+}$ ions are located in "A" and "D" sites, while "C" is occupied by $Y^{3+}$ ions. During the vacuum sintering, chromium incorporated into YAG as $Cr^{3+}$. Air annealing led to recharge the part of $Cr^{3+}$ ions into tetravalent state. It is worth to notice that $Cr^{3+}$ ions occupy only octahedral sites. Furthermore, the octahedrally coordinated $Cr^{4+}$ ions exchange their positions with tetrahedrally coordinated $Al^{3+}$ ions and as a result, the portion of $Cr^{4+}$ ions occupy tetrahedral sites. The following chromium oxidation chain can be proposed: $Cr_A^{3+} \rightarrow Cr_A^{4+} \rightarrow Cr_D^{4+}$, where $Cr_A^{3+}, Cr_A^{4+}, Cr_D^{4+}$ represent octahedrally coordinated $Cr^{3+}$ ions and octahedrally/tetrahedrally coordinated $Cr^{4+}$ ions, respectively. More detailed explanation of chromium valence transformation was reported elsewhere [12,17].

The estimation of $Cr^{4+}$ content in Cr-doped garnets is an extremely important task, which is, however, not trivial. We propose an approach based on the optical spectroscopy methods to estimate the $Cr^{4+}$ concentration in Cr:YAG ceramics using the survey absorption spectra. The Smakula–Dexter formula can be used for this purpose, however, uncertainties in the values of oscillator strengths for $Cr^{4+}$ absorption bands and, moreover, uncertainties in the deconvolution of $Cr^{4+}$ absorption spectra make it difficult to calculate $Cr^{4+}$ concentration with high accuracy. In fact, the oscillator strengths for $Cr^{4+}$ ions absorption bands were calculated based on the energy fluence-dependent transmission at 1064 nm, which implies uncertainty in the obtained results. In order to understand the origin of this uncertainty, a step-by-step calculation procedure is described below.

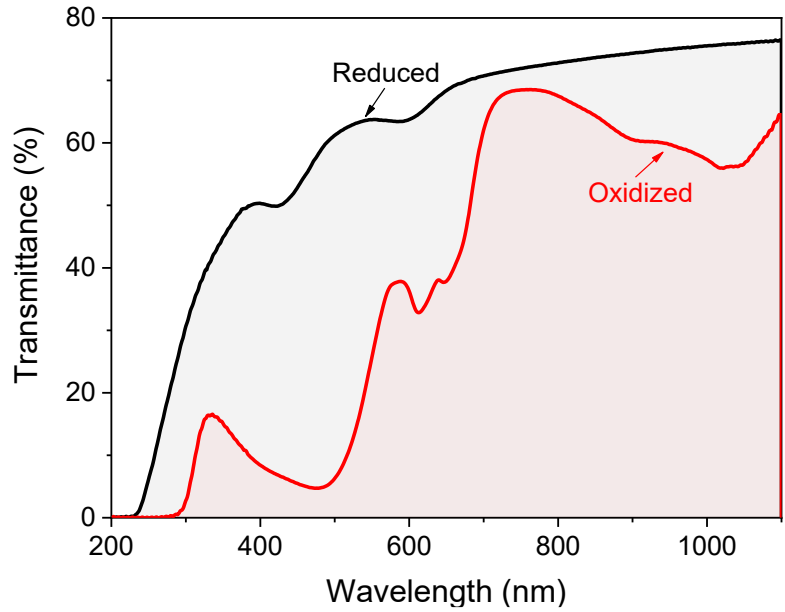

Fig. 2: Optical in-line transmittance of fully reduced (black line) and fully oxidized (red line) $Cr^{4+}$:YAG ceramics.

At first, the transmission and absorption spectra of both as-grown Cr:YAG ceramic (fully reduced) and after air annealing (fully oxidized) were measured (Figs. 2, S1). In-line transmission of the reduced samples was 76% at 1100 nm. Measured absorption spectra show that all chromium ions are in trivalent state after vacuum sintering [23-24]. The characteristics of the $Cr^{3+}$ absorption bands are collected in Table S1. $Cr^{3+} \rightarrow Cr^{4+}$ transition was stimulated by air annealing of the samples at 1400 °C for 15 hours, which leads to a decrease in transmittance due to formation of the $Cr^{4+}$ ions (Fig. 2). An increase in the absorption at 480 nm and 1030 nm is caused by $Cr_A^{4+}$ and $Cr_D^{4+}$ ions respectively (Fig. S1).

In the second step, the concentration of $Cr^{4+}$ ions was estimated form absorption spectra [12]. The difference between absorption spectra of fully oxidized and fully reduced $Cr^{4+}$,$Ca^{2+}$:YAG ceramics was attributed to the absorption of $Cr^{4+}$ ions. The effect of $Cr^{3+}$ concentration decrease on the spectrum should be noted. However, due to the fact that only up to 25% of trivalent chromium ions become tetravalent [17], and the absorption cross-section of $Cr^{4+}$ is several times more than the one of $Cr^{3+}$ [16], the effect of the change of $Cr^{3+}$ concentration on the absorption spectra is limited.

Fig. 3 and Fig. 4 show the calculated spectra obtained as a difference between absorption of fully oxidized and fully reduced spectra of $Cr^{4+}$,$Ca^{2+}$:YAG ceramics. The calculated spectra were fitted by the FITYK 0.9.8 program [25] to deconvolute the individual peaks into log-normal functions [26]. The log-normal curve (Eq. 1) is an asymmetric Gaussian curve which shape is given by four parameters, $\acute{\upsilon}_0$, (position) H, (height) $\delta_{FWHM}$ (half-width) and p (skewness)

$$\varepsilon(\acute{\upsilon}) = H \cdot exp\left\{\frac{-\ln 2}{(\ln p)^2}\left[\ln\left(\frac{\acute{\upsilon}-\acute{\upsilon}_0}{\delta_{FWHM}}\frac{(p^2-1)}{p}+1\right)\right]^2\right\} \qquad (1)$$

The skewness is defined as $(\acute{\upsilon}_0 - \acute{\upsilon}_v)/(\acute{\upsilon}_r - \acute{\upsilon}_0)$ where $\acute{\upsilon}_r$ and $\acute{\upsilon}_v$ are the value of $\acute{\upsilon}$ at $\varepsilon_0/2$ on the lower energy side (red) and the higher energy side (blue), respectively. The log-normal curve become the Gaussian one as the skewness is 1 [27]. Absorption and emission bands are normally characterized by a skewness parameter p > 1 and p < 1 respectively.

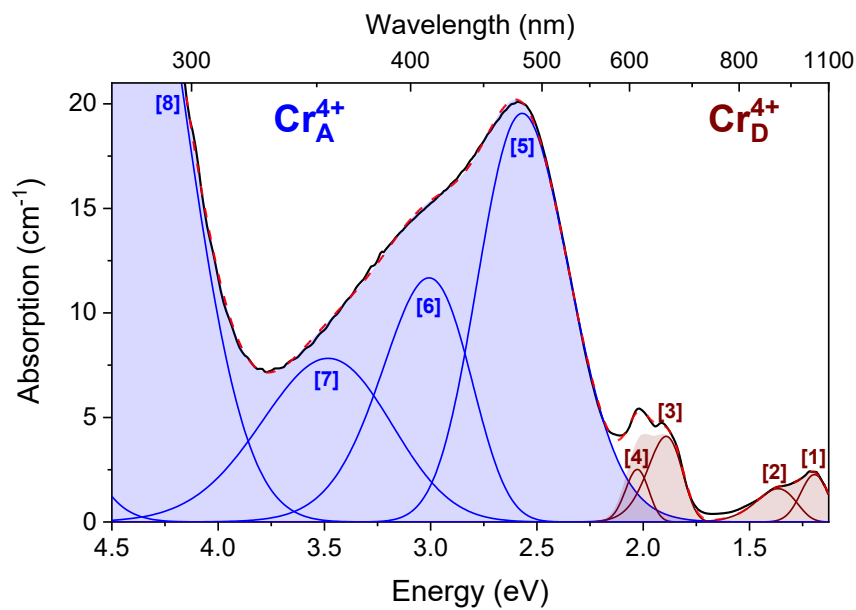

Fig. 3. Decomposition of the absorption spectrum of the $Cr^{4+}$:YAG ceramics after vacuum sintering at 1750 °C (15 h) followed by air annealing at 1450 °C (15 h) using an approach proposed by Feldman et. al. [12]. Blue field corresponds absorption of octahedral $Cr^{4+}$, light brown – tetrahedral $Cr^{4+}$.

A major problem in the fitting of the $Cr^{4+}$ absorption spectra is the necessity to adjust four parameters of the log-normal curves for each absorption band. All four parameters can be measured directly in case of well-separated bands, but for overlapping bands, as in our case, no well-defined set of parameters can be obtained from the data. Consequently, the parameters of $Cr^{4+}$ absorption bands are difficult to find, moreover, the number of peaks and their positions are unknown. It is a common problem for 3d luminescent ions, for which the impact of ligand dynamics on their transitions is considerable and as result, the absorption bands are fairly broadened.

The $Cr^{4+}$ absorption spectra were fitted by the log-normal function based on two different sets of absorption bands proposed by Feldman et. al. [12] (Fig. 3) and Ubizskii et. al. [27] (Fig. 4), who have used the modified Lorentzian function and Gaussian function. These band sets differ in number and the parameters of peaks. Both of them were successfully applied in other works for different $Cr^{4+}$:YAG materials [7,12,17,18,19,28,29,30]. The peak parameters for Feldman and Ubizskii approaches are collected in Table S2 and Table S3, respectively.

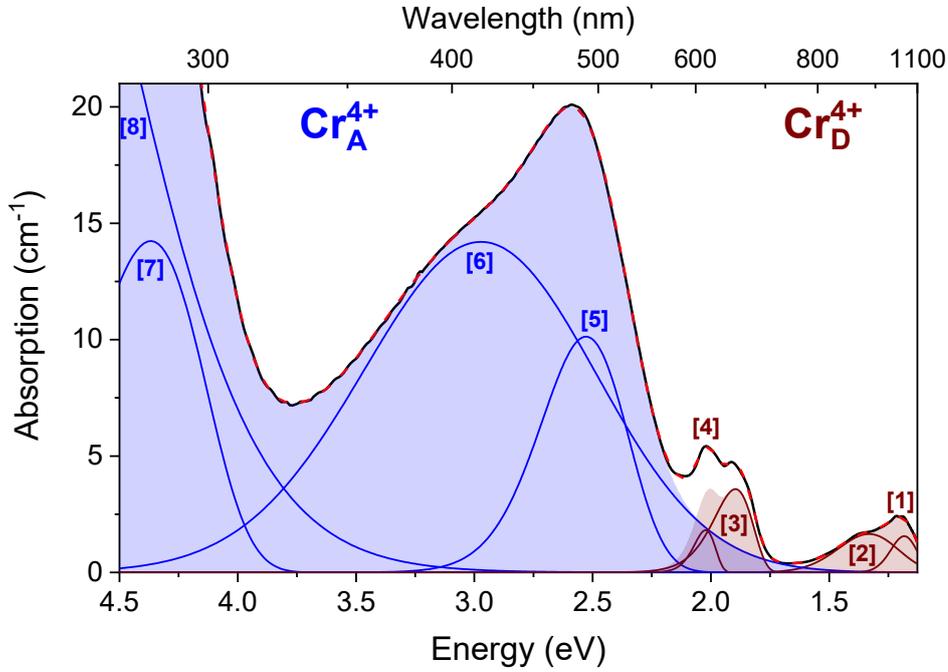

Fig. 4. Decomposition of the absorption spectrum of the $Cr^{4+}$:YAG ceramics after vacuum sintering at 1750 °C (15 h) followed by air annealing at 1450 °C (15 h) using an approach proposed by Ubizskii et. al. [27]. Blue field corresponds absorption of octahedral $Cr^{4+}$, light brown – tetrahedral $Cr^{4+}$.

The concentrations of $2 \cdot 10^{18}$ ions/cm³ and $1.1 \cdot 10^{18}$ ions/cm³ for $Cr_A^{4+}$ and $Cr_D^{4+}$ ions, respectively, were found using the Smakula-Dexter formula (Eq. 2) [31].

$$n = 0.87 \cdot 10^{17} \frac{a_0}{(a_0^2+2)^2} \cdot \frac{\delta_{FWHM}}{f} H \qquad (2)$$

where n is the concentration of absorbing centers, $a_0$ is the refractive index of the crystal in the maximum of the absorption band (the value was taken from [32]), and f is the oscillator strength. Both f ($2.3 \cdot 10^{-3}$ and $2.2 \cdot 10^{-2}$ for band #1 and #5, respectively) and $\delta_{FWHM}$ (0.21 eV and 0.55 eV for band #1 and #5, respectively) parameters were taken from literature [12] for unifying the calculated value of $Cr^{4+}$ concentration in our works [7,17,18,20,21], while the H was taken from Table S2. It should be noted that the oscillator strength calculated earlier by Feldman et. al. using defined peak parameters [12] cannot be used for the Ubizskii approach.

It was previously shown [11,16,33], that $Cr^{4+}$ content in YAG materials can be successfully estimated using properly calculated ground-state and excited-state absorption cross-sections of $Cr_D^{4+}$ ions. We have adopted this calculation to our study to verify the obtained values. The main feature of $Cr_D^{4+}$ ions is the absorption saturation that caused decrease in absorption intensity with rising the excitation power. In other way, the sample transparency increases with increase of the excitation source power at 1 µm. The transmittance as a function of input energy fluence obtained for Cr,Ca:YAG ceramic sample is shown in Fig. 5.

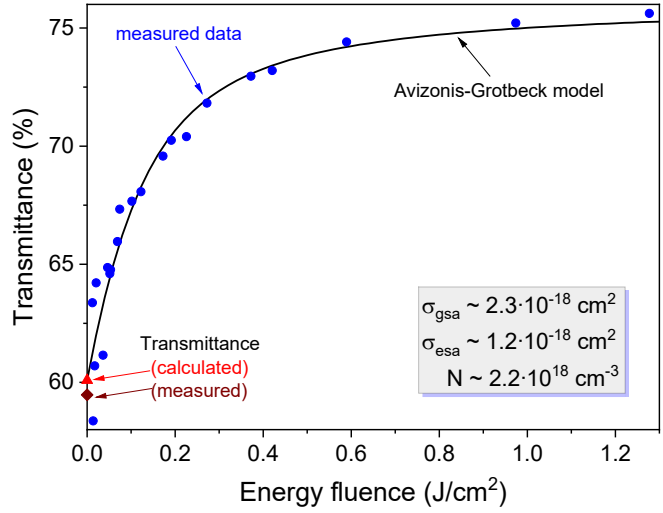

Fig. 5: Transmittance at 1064 nm (blue circles) as a function of the input energy fluence for the Cr,Ca:YAG ceramic sample. The pump pulse duration was 7 ns (FWHM) in the single shot mode. Solid curve represent best fitting by means of Avizonis-Grotbeck model (Eq. 3).

The spectroscopic parameters, including concentration of $Cr_D^{4+}$ ions in $Cr^{4+}$:YAG ceramics were calculated using the Avizonis-Grotbeck theory [11]. The following differential equation describes evolution of the energy density E(z) of incident radiation along the sample with thickness $d$ in $z$ direction [11,33]. Since the $^3B_1(^3A_2) \rightarrow {}^3A_2(^3T_1)$ relaxation time (~3.4 µs) is considerably longer than the experimental pulse duration (7 ns), the results of the energy fluence-dependent transmission at 1064 nm were analyzed with the following equation, which accounts for the excited state absorption of the absorbing media:

$$\frac{dE(z)}{dz} = -\hbar\omega N \left\{\left(1 - \frac{\sigma_{esa}}{\sigma_{gsa}}\right)\left[1 - exp\left(-\frac{\sigma_{gsa}E(z)}{\hbar\omega}\right)\right] + \frac{\sigma_{esa}E(z)}{\hbar\omega}\right\} - \alpha E(z) \qquad (3)$$

where $N$ is the concentration of active ions, $\sigma_{gsa}$ and $\sigma_{esa}$ are the ground state and excited state absorption cross-sections, respectively, $\hbar\omega$ is the photon energy, and $\sigma$ are the unsaturated losses. Equation (3) was solved to give the best fit for the experimental data of transmission as a function of energy fluence. The best agreement was obtained for $\sigma_{gsa} = 2.32 \times 10^{-18}$ cm$^2$ and $\sigma_{esa} = 1.22 \times 10^{-18}$ cm$^2$, $N = 2.2 \cdot 10^{18}$ ion/cm$^3$, unsaturated losses ~ 0.1 cm$^{-1}$.

Table 1. Measured absorption coefficients a, densities N, absorption cross-section parameters $\sigma_{gsa}$ and $\sigma_{esa}$, and oscillator strengths $f^*$ of the 1030 nm absorption band (Fig. 3) of some $Cr^{4+}$:YAG materials. The oscillator strengths were calculated using reported N and a values.

| Compounds | Type | $\sigma_{gsa}$, $10^{-18}$ cm² | $\sigma_{esa}$ | $\sigma_{esa}/\sigma_{gsa}$ | N, $Cr_D^{4+}$ $10^{18}$ cm⁻³ | a cm⁻¹ | $f^*$ $10^{-3}$ | Ref. |
|---|---|---|---|---|---|---|---|---|
| $Cr^{4+}$:YAG | Ceramic | 2.3 | 1.2 | 0.52 | 2.1 | 2.2 | 1.1 | This work |
| $Cr^{4+}$:YAG | Single crystal | 2.11 | 0.2 | 0.08 | 2.4 | 5.01 | 2.2 | [16] |
| $Cr^{4+}$:YAG | Single crystal | 3.93 | 0.5 | 0.14 | 0.3 | 1.2 | 4.2 | [16] |
| $Cr^{4+}$:YAG | Single crystal | 4.59 | 1.1 | 0.24 | 0.3 | 1.21 | 4.9 | [16] |
| $Cr^{4+}$:YAG | Single crystal | 2.43 | 0.4 | 0.18 | 0.4 | 1.01 | 2.6 | [16] |
| $Cr^{4+}$:YAG | Single crystal | 4.6 | 0.7 | 0.14 | 0.2 | 1.03 | 4.9 | [16] |
| $Cr^{4+}$:YAG | Single crystal | 7 | 2 | 0.29 | 0.5 | | | [13] |
| $Cr^{4+}$:YAG | Single crystal | 0.8 | 0.1 | 0.13 | 1 | | 2.3 | [12] |
| $Cr^{4+}$:YAG | Single crystal | 5.7 | 0.8 | 0.14 | | | | [15] |
| $Cr^{4+}$:YAG | Single crystal | 3.2 | 0.6 | 0.19 | 0.19 | | | [14] |

The concentration of $Cr_D^{4+}$ ions calculated from Avizonis-Grotbeck theory was 2.2·10¹⁸ ion/cm⁻³, which is twice the value obtained from the optical absorption spectra. This difference can be explained by the uncertainty in the calculated value of the spectroscopic parameters of $Cr_D^{4+}$. As a result, reported parameters, including $\sigma_{gsa}$ and $\sigma_{esa}$, differ in one order of magnitude from the ones reported previously [13-16], see Table 1. In addition, the oscillator strengths also can be determined with some accuracy (more details see in Discussion) leading to the difference in the reported values of $Cr^{4+}$ concentrations in $Cr^{4+}$:YAG materials with close values of $Cr^{4+}$ absorption. According to our knowledge, there were no attempts to determine the confidential range for calculated concentrations of $Cr^{4+}$ ions. It should be noted that the $Cr_D^{4+}/Cr_A^{4+}$ ratio can be calculated with high precision in contrast to the total concentration of $Cr^{4+}$ ions as the concentration of $Cr_A^{4+}$ ions can be calculated form the decrease in the absorption of $Cr_A^{4+}$ ions during $Cr_A^{4+} \rightarrow Cr_D^{4+}$ transition [12].

4. Discussion

The concentration of $Cr^{4+}$ ions has been calculated from power dependence of optical transitions of $Cr^{4+}$-doped garnets under pulsed 1 μm laser as shown in Fig. 5. However, the obtained values can differ from the actual ones due to sufficient limitations of the method. Moreover, this method is complicated and requires special equipment, as well as knowledge of the Avizonis-Grotbeck model (eq. 3) that limits its applicability. On the other hand, the Smakula-Dexter formula (eq. 2) requires only optical absorption spectra and therefore can be easily used. At the same time, the Smakula-Dexter formula uses the parameters of the absorption peaks and the oscillator strengths, that are obtained with some errors. So, the fitting model should be used. This model is based on the number and positions of the peaks proposed by Feldman (Fig. 3) or Ubizskii (Fig. 4). A major problem in the fitting of the $Cr^{4+}$ absorption spectra is the necessity to adjust four parameters for each absorption band. For well-separated bands, all four parameters can be measured directly, but for overlapping bands, as in our case, no unique set of parameters can be extracted from the data. Consequently, the $Cr^{4+}$ absorption band parameters are difficult to find out.

The calculated values of oscillator strength of the band at 1030 nm are in the range from $1.1 \cdot 10^{-3}$ to $5 \cdot 10^{-3}$ implying the uncertainty in the calculated values of $C(Cr_D^{4+})$. Originally, the oscillator strengths for $Cr^{4+}$ absorption bands were calculated by the method proposed by Feldman et. al. from the spectroscopic properties of $Cr^{4+}$ ions using the power dependence of optical transmittance of $Cr^{4+}$:YAG single crystals (1064 nm), see Fig. 5. However, this method gives wide spread of obtained parameters, and so, uncertainties in oscillator strength values. Table 1 collects the calculated values of the oscillator strengths for 1030 nm absorption bands of $Cr_D^{4+}$ ions using earlier reported concentration of $C(Cr_D^{4+})$ and absorption coefficient at 1030 nm. The calculated oscillator strengths range from $1.1 \cdot 10^{-3}$ to $5 \cdot 10^{-3}$ (Table 1).

Described earlier problems can be efficiently eliminated by some simplifications to the Smakula-Dexter formula. The results obtained in this work allow us to propose the possible way to calculate the concentration of $Cr^{4+}$ ions in both octahedral and tetrahedral positions based on Smakula-Dexter formula (Eq. 2). Beforehand, we should suggest that $Cr^{4+}$ absorption bands differ in the various $Cr^{4+}$:YAG materials only by the value of absorption coefficient at the band maximum while other parameters (peak positions, asymmetry and FWHM) are the same. This simplification eliminates the problem with the fitting of the $Cr^{4+}$ absorption spectrum, since the Smakula-Dexter formula can be reduced to the following equation: $n = C \cdot \varepsilon \cdot H_0$, where $C = 0.87 \cdot 10^{17} (a_0/(a_0^2 + 2)^2) \cdot (\delta_{FWHM}/f)$ is constant, $\varepsilon$ is coefficient of proportionality between the absorption coefficient at the band maximum and total absorption at the same wavelength, and $H_0$ is total absorption coefficient for $Cr^{4+}$ ions (at 480 nm and 1030 nm for $Cr_A^{4+}$ ions and $Cr_D^{4+}$ ions, respectively), which is calculated as the difference between the ones obtained for fully oxidized and fully reduced samples. As a result, the concentration of $Cr^{4+}$ ions in both octahedral and tetrahedral positions can be calculated using the absorption coefficients of $Cr^{4+}$ ions at 480 nm and 1030 nm, respectively.

The concentration of $Cr^{4+}$ ions in octahedral and tetrahedral sites can be calculated by the following equation:

$$n_A = 1.2 \cdot 10^{17} H_{480\,nm} \; ; \quad n_D = 5.4 \cdot 10^{17} H_{1030\,nm} \qquad (4)$$

where $n_A$, cm$^{-3}$ and $n_D$, cm$^{-3}$ are concentrations of $Cr_A^{4+}$ ions and $Cr_D^{4+}$ ions, respectively. This formula can be applied for other $Cr^{4+}$-doped garnets with similar shapes of optical absorption spectra due to similarities in the spectroscopic parameters (Table S4). These formulas have been obtained by dividing the calculated values of $C(Cr_A^{4+})$ ($2 \cdot 10^{18}$ ions/cm$^3$) and $C(Cr_D^{4+})$ ($1.1 \cdot 10^{18}$ ions/cm$^3$) by the measured absorption coefficient at 480 nm (20.1 cm$^{-1}$) and 1030 nm (2.4 cm$^{-1}$), respectively. It should be noted that the concentration of Cr$^{4+}$ ions was calculated by Smakula-Dexter formula (eq. 2) using the oscillator strength reported by Feldman et. al ($2.3 \cdot 10^{-3}$), while its actual value is in the range from $1.1 \cdot 10^{-3}$ to $5 \cdot 10^{-3}$. So, the actual concentration of Cr$^{4+}$ ions can be from half up to the twice of value obtained by eq. 4. Therefore, the actual concentration $n_0$ is in the range $n/2 < n_0 < 2 \cdot n$ where n is calculated value by eq. 4.

The given formula and confidence range related to the C(Cr$^{4+}$) affect the findings related to concentration of divalent ions required for charge compensation reported earlier [7,17-20]. Beforehand was shown that the concentration of Ca$^{2+}$ ions in the grain volume of Ca,Cr:YAG ceramics was up to ~$1.1 \cdot 10^{19}$ cm$^{-3}$, while the calculated concentration of Cr$^{4+}$ ions was ~$0.6 \cdot 10^{19}$ cm$^{-3}$, which is one half of Ca$^{2+}$ concentration. Applying the confidential range ($n/2 < n_0 < 2 \cdot n$), the actual value of Cr$^{4+}$ concentration can be found in the range from ~$0.3 \cdot 10^{19}$ cm$^{-3}$ to $1.1 \cdot 10^{19}$ cm$^{-3}$, which means that from 25% up to 100% of Ca$^{2+}$ ions are involved in chromium valence transformation. At the same time, according to the current point of view, only a small fraction of Ca$^{2+}$ ions contribute to chromium valence transformation [7,12]. Probably, that the concentration of Cr$^{4+}$ ions can be limited by solubility of Ca$^{2+}$ ions in grain volume of garnet materials. Unfortunately, the existing model lacks of details explaining all features of Cr$^{4+}$ ions formation. The present paper is the first step to understand the Cr$^{4+}$ formation providing a useful tool to verify the range of actual Cr$^{4+}$ concentrations.

**Conclusion**

In summary, the possible approach for calculation of Cr$^{4+}$ concentration in Cr$^{4+}$:YAG transparent materials was proposed. Convenient optical spectroscopy methods make possible to estimate the Cr$^{4+}$ concentration in Cr:YAG ceramics using survey absorption spectra. The Smakula–Dexter formula can be used for this purpose, however, uncertainty in the value of oscillator strength for Cr$^{4+}$ absorption bands, and, moreover, uncertainty in the deconvolution of Cr$^{4+}$ absorption spectra make it difficult to calculate the Cr$^{4+}$ ions concentration with high accuracy. The following formulas were proposed for calculation of Cr$^{4+}$ concentration: $n_A = 1.2 \cdot 10^{17} H_{480}$, $n_D = 5.4 \cdot 10^{17} H_{1030}$, where $n_A$, cm$^{-3}$ and $n_D$, cm$^{-3}$ are calculated concentrations of octahedrally and tetrahedrally coordinated Cr$^{4+}$ ions, respectively, $H_{480}$ and $H_{1030}$ are absorption coefficients of Cr$^{4+}$ ions at 480 nm and 1030 nm, respectively. The actual concentration $n_0$ is in the range $n/2 < n_0 < 2 \cdot n$, where n is calculated value.

**Acknowledge**

The authors are grateful to Dr. A.G. Doroshenko and Dr. S.V. Parkhomenko for their help in sintering of the sample.

**Supplementary files**

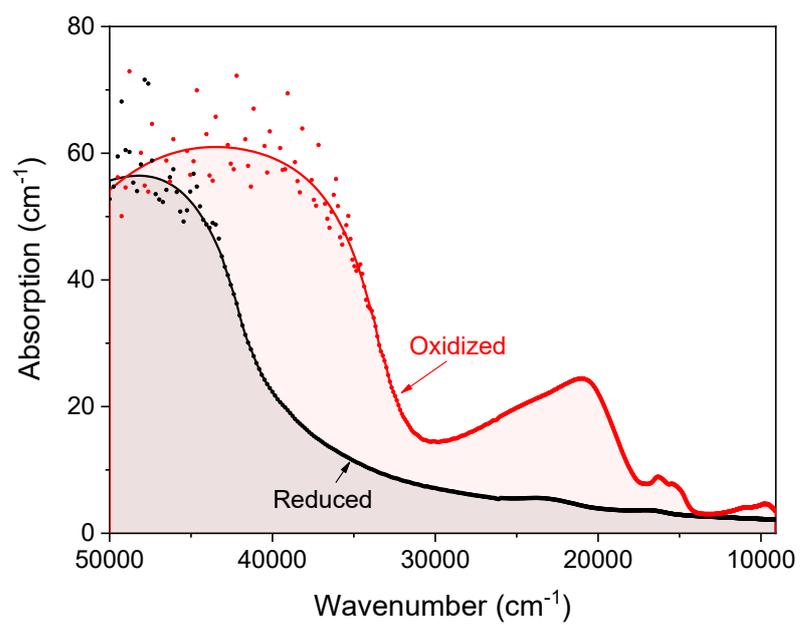

Fig. S1: Optical absorption spectra of fully reduced (black line) and fully oxidized (red line) $Cr^{4+}$:YAG ceramics.

Table S1: Characteristics of the absorption bands of $Cr^{3+}$ ions for the fully reduced $Cr^{4+}$:YAG ceramics: the assignment, $\lambda_p$ - peak positions. $\delta_{FWHM}$ - full width at half maximum, $H$ - height, p - asymmetry coefficient.

| No. | Assignment | $\lambda_p$ [nm] | | | $\delta_{FWHM}$ | | H | σ | p |
| --- | --- | --- | --- | --- | --- | --- | --- | --- | --- |
| | | [nm] | [eV] | [cm$^{-1}$] | [eV] | [cm$^{-1}$] | [cm$^{-1}$] | cm$^2$ | |
| 1 | $^4A_{2g} \rightarrow {}^4T_{1g}$ | 434 | 2.86 | 23061 | 0.41 | 3284 | 0.90 | 46 10$^{-21}$ | 0.87 |
| 2 | $^4A_{2g} \rightarrow {}^4T_{2g}$ | 597 | 2.08 | 16738 | 0.28 | 2232 | 0.45 | 23 10$^{-21}$ | 1.24 |

Table S2: Characteristics of the absorption bands of $Cr^{4+}$ ions in $Cr^{4+}$,$Ca^{2+}$:YAG ceramic (Fig. 3): the assignment, $\lambda_p$ - peak positions. $\delta_{FWHM}$ - full width at half maximum, $H$ - height, p - asymmetry coefficient.

| No. | Assignment | $\lambda_p$ [nm] | | | $\delta_{FWHM}$ | | H | p |
| --- | --- | --- | --- | --- | --- | --- | --- | --- |
| | | [nm] | [eV] | [cm$^{-1}$] | [eV] | [cm$^{-1}$] | [cm$^{-1}$] | |
| 1 | $^3B_1(^3A_2) \rightarrow {}^3A_2(^3T_1)+[v]$ | 1039 | 1.19 | 9627 | 0.15 | 1235 | 2.1 | 1.03 |
| 2 | $^3B_1(^3A_2) \rightarrow {}^3E(^3T_2)$ | 908 | 1.37 | 11014 | 0.22 | 1798 | 1.5 | 1.23 |
| 3 | $^3B_1(^3A_2) \rightarrow {}^3E(^3T_1)$ | 655 | 1.89 | 15257 | 0.19 | 1554 | 4.2 | 1.20 |
| 4 | $^3B_1(^3A_2) \rightarrow {}^3E(^3T_1)+[v]$ | 612 | 2.03 | 16351 | 0.13 | 1039 | 2.0 | 1.05 |
| 5 | $^3T_1 \rightarrow {}^3T_2$ | 483 | 2.57 | 20721 | 0.50 | 4065 | 14.3 | 0.91 |
| 6 | Not assigned | 412 | 3.01 | 24255 | 0.49 | 3982 | 10.2 | 1.17 |
| 7 | Not assigned | 356 | 3.48 | 28083 | 0.72 | 5797 | 3.97 | 1.07 |
| 8 | $^3T_1 \rightarrow {}^3T_1$ | 274 | 4.53 | 36514 | 0.83 | 6719 | 30.9 | 1.13 |

Table S3: Characteristics of the absorption bands of $Cr^{4+}$ ions in $Cr^{4+},Ca^{2+}$:YAG ceramic (Fig. 4): the assignment, $\lambda_p$ - peak positions. $\delta_{FWHM}$ - full width at half maximum, $H$ - height, p - asymmetry coefficient.

| No. | Assignment | $\lambda_p$ [nm] | | | $\delta_{FWHM}$ | | H | p |
|---|---|---|---|---|---|---|---|---|
| | | [nm] | [eV] | [cm$^{-1}$] | [eV] | [cm$^{-1}$] | [cm$^{-1}$] | |
| 1 | $^3B_1(^3A_2) \rightarrow {}^3A_2(^3T_1)$+[v] | 1049 | 1.18 | 9533 | 0.13 | 1080 | 1.52 | 0.92 |
| 2 | $^3B_1(^3A_2) \rightarrow {}^3E(^3T_2)$ | 937 | 1.32 | 10668 | 0.30 | 2391 | 1.74 | 1.20 |
| 3 | $^3B_1(^3A_2) \rightarrow {}^3E(^3T_1)$ | 656 | 1.89 | 15249 | 0.17 | 1370 | 3.61 | 1.20 |
| 4 | $^3B_1(^3A_2) \rightarrow {}^3E(^3T_1)$+[v] | 613 | 2.02 | 16314 | 0.12 | 981 | 2.36 | 1.23 |
| 5 | $^3T_1 \rightarrow {}^3T_2$ | 491 | 2.53 | 20379 | 0.42 | 3412 | 9.61 | 1.14 |
| 6 | Not assigned | 420 | 2.95 | 23811 | 1.17 | 9474 | 14.50 | 1.08 |
| 7 | Not assigned | 284 | 4.37 | 35224 | 0.57 | 4620 | 14.61 | 1.21 |
| 8 | $^3T_1 \rightarrow {}^3T_1$ | 254 | 4.89 | 39424 | 1.00 | 8058 | 35.68 | 0.83 |

Table S4. Summary of measured density N, and absorption cross-section parameters $\sigma_{gsa}$ and $\sigma_{esa}$ of some chromium-doped materials

| Compounds | Type | $\sigma_{gsa}$, $\sigma_{esa}$ $10^{-18}$ cm$^2$ | | $\sigma_{esa}/\sigma_{gsa}$ | N, $Cr_D^{4+}$ $10^{18}$ cm$^{-3}$ | Ref. |
|---|---|---|---|---|---|---|
| $Cr^{4+}$:LuAG | Single crystal | 2.43 | 0.3 | 0.11 | 0.3 | [16] |
| $Cr^{4+}$:LuAG | Single crystal | 2.89 | 0.6 | 0.20 | 0.3 | [16] |
| $Cr^{4+}$:LuAG | Single crystal | 3.25 | 0.2 | 0.07 | 0.2 | [16] |
| $Cr^{4+}$:LuAG | Single crystal | 1.43 | 0.2 | 0.16 | 0.5 | [16] |
| $Cr^{4+}$:LuAG | Single crystal | 1.96 | 0.1 | 0.08 | 0.4 | [16] |
| $Cr^{4+}$:GGG | Single crystal | 5.8 | 1.3 | 0.22 | 0.07 | [13] |
| $Cr^{4+}$:YSGG | Single crystal | 4.5 | 0.4 | 0.09 | 0.8 | [13] |
| $Cr^{4+}$:YSGG | Single crystal | 4.6 | 0.4 | 0.09 | 0.7 | [13] |
| $Cr^{4+}$:LuAG | Single crystal | 1.1 | 0.045 | 0.04 | 2.6 | [13] |